\documentclass{aastex}
\usepackage{emulateapj5}
\slugcomment{Accepted in ApJ July 2001}
\usepackage{graphicx}

\begin{document}

\title{Star Formation in M51 Triggered by Galaxy Interaction}
\author{
 Nikola, T. \altaffilmark{1}, 
 Geis, N. \altaffilmark{2}, 
 Herrmann, F. \altaffilmark{2}, 
 Madden, S.C. \altaffilmark{3}, 
 Poglitsch, A. \altaffilmark{2}, 
 Stacey, G.J. \altaffilmark{1}, \& 
 Townes, C.H. \altaffilmark{4}
}

\altaffiltext{1}{Department of Astronomy, Cornell University, 
 Ithaca, NY 14853}
\altaffiltext{2}{Max-Planck-Institut f{\"u}r extraterrestrische Physik, 
 Garching, Germany}
\altaffiltext{3}{CE-Saclay, Service d'Astrophysique, 91191 Gif-sur-Yvette, 
 France}
\altaffiltext{4}{Department of Physics, University of California, Berkeley, 
 CA 94720}

\begin{abstract}

We have mapped the inner $360''$ regions of M51 in the 158~$\mu$m 
[C~{\small II}] line at $55''$ spatial resolution using the Far-infrared 
Imaging Fabry-Perot Interferometer (FIFI) on the Kuiper Airborne 
Observatory (KAO).
The emission is peaked at the nucleus, but is detectable over the entire 
region mapped, which covers much of the optical disk of the galaxy.
There are also two strong secondary peaks at $\sim$~43\% to 70\% of the 
nuclear value located roughly $120''$ to the north-east, and south-west
of the nucleus.
These secondary peaks are at the same distance from the nucleus as the 
corotation radius of the density wave pattern.
The density wave also terminates at this location, and the outlying spiral 
structure is attributed to material clumping due to the interaction 
between M51 and NGC~5195.
This orbit crowding results in cloud-cloud collisions, stimulating star
formation, that we see as enhanced [C~{\small II}] line emission.
The [C~{\small II}] emission at the peaks originates mainly from 
photodissociation regions (PDRs) formed on the surfaces of molecular 
clouds that are exposed to OB starlight, so that these [C~{\small II}]
peaks trace star formation peaks in M51.
The total mass of [C~{\small II}] emitting photodissociated gas is 
$\sim 2.6 \times 10^{8} M_{\odot}$, or about 2\% of the molecular gas
as estimated from its CO(1$\to$0) line emission.
At the peak [C~{\small II}] positions, the PDR gas mass to total gas mass 
fraction is somewhat higher, 3$\to$17\%, and
at the secondary peaks the mass fraction of the [C~{\small II}] 
emitting photodissociated gas can be as high as 72\% of the molecular mass. 
Using PDR models, we estimate the far-UV field intensities are a few 100
times the local Galactic interstellar radiation field, similar to that 
found near OB star forming giant molecular clouds in the Milky Way.
The density solution is degenerate, with both a low ($n \sim 10^{2} - 
10^{3}$~cm$^{-3}$), and a high ($n \sim 10^{3} - 10^{6}$~cm$^{-3}$),
density solution.
Our analysis shows that a substantial amount of the observed 
[C~{\small II}] emission from the galaxy as a whole can arise from the 
ionized medium, and that the contribution from the cold neutral medium 
(CNM) is not negligible.
At the [C~{\small II}] peaks, probably $\sim$~7 -- 36\% of the 
[C~{\small II}] emission arises from the CNM, while north-west of 
the nucleus, most of the observed emission may arise 
from the CNM. 

\end{abstract}

\keywords{galaxies: interaction, individual(M51) --infrared: galaxies}

\section{Introduction}

We are carrying out a study of the properties of the interstellar medium 
(ISM) in galaxies to understand the effects of the interplay between the 
global conditions of a galaxy and its star formation activity.
The global properties of a galaxy most likely set the preconditions for
star formation and also trigger the star formation.
By ``global'' we are refering to the conditions in the spiral arms, interarm
medium, nuclei, or individual regions in the few hundred parsec range.
Interacting galaxies are excellent laboratories for this investigation
because the interaction can speed up processes (e.g. increase mass flow 
rates)
or enhance physical conditions (e.g. increase densities) in the 
interstellar medium.
Since every galaxy is unique in some sense it is important to carry
out case studies of individual interacting galaxies.
Here we have chosen the galaxy M51 to investigate its ISM and its star 
formation activity.

The galaxy M51 (NGC~5194), also called the whirlpool galaxy, is a 
grand-design spiral of Hubble type Sbc.
It has an inclination angle of $\approx 20^{\circ}$ (Tully 1974 b) and 
thus it is seen almost face on.
The total spatial extent of M51, as seen in the visible, is about $7'$.
M51 is interacting with its companion NGC~5195 which is $4.5'$ to the north.
Due to its proximity (9.6~Mpc, Sandage \& Tammann 1975) and its face on 
appearance, M51 is one of the best studied interacting galaxies.
Despite this fact some aspects of this galaxy remain a puzzle.

The star formation rate of M51 is not spectacular.
It is quite normal for an isolated Sbc-Sc galaxy (Kennicutt 1998).
However, M51 might have been a Sb galaxy prior to the 
interaction with NGC~5195 and evolved to a Sbc-Sc late-type galaxy (Tully 
1974 c; Kennicutt 1998).
The galaxy interaction is also believed to be responsible for the current 
density wave in M51 enhancing the inner spiral arm structure and 
triggering star formation in the spiral arms.
The density wave pattern, however, terminates at corotation and the spiral
arms beyond corotation are attributed to material clumping induced directly
by the galaxy interaction (Tully 1974 c).
It is clear from the arrangement of the ionized gas
(Tully 1974 a,b,c; van der Hulst et al.\ 1988; Tilanus \& Allen 1991),
molecular gas
(Garc\' \i a-Burillo, Gu\'elin, \& Cernicharo 1993; Lo et al.\ 1987; 
Rand \& Kulkarni 1990; Vogel, Kulkarni, \& Scoville 1988), atomic gas 
(Tilanus \& Allen 1989, 1991; Rots et al.\ 1990),
and dust in the spiral arms that star formation in the arms is 
due to a density wave.
However, the crowding of H~{\small II} regions in the north-east and the
south-west of M51, as seen for example in the H$\alpha$ map of van der Hulst 
et al.\ (1988), 
may not be caused by the density wave but directly linked to the 
galaxy interaction.
In this case the galaxy interaction has strong and direct control over the 
location and the strength of the star formation in M51.

To investigate the star formation activity, we observed M51 in 
the 158~$\mu$m [C~{\small II}] fine-structure line. 
The [C~{\small II}] line emission from galaxies arises predominantly in the 
warm, dense, photodissociated  surfaces of molecular clouds exposed to 
starlight from nearby OB stars (e.g. Stacey et al.\ 1991). 
These regions, commonly called photodissociation regions (PDRs) are heated
predominantly by photoelectric ejection of energetic electrons from grains,
and cooled by far-infrared (FIR) fine structure line radiation from 
O$^{\circ}$, and C$^{+}$.
Typically, the strongest cooling line is the [C~{\small II}] line, but the
[O~{\small I}] 63~$\mu$m line is often roughly the same strength in 
starburst nuclei.
The [C~{\small II}] line emission from moderate velocity shocks is much 
weaker than that from PDRs (Hollenbach and McKee 1989).

\section{Observation}
\label{observation}

The observation of the [C~{\small II}] 
$^{2}P_{3/2}$$\rightarrow$$^{2}P_{1/2}$ fine structure line at 
157.7409~$\mu$m was carried out with the MPE/UCB Far-infrared Imaging
Fabry-Perot Interferometer, FIFI (Poglitsch et al.\ 1991; Stacey  
et al.\ 1992), on board the Kuiper Airborne Observatory in June 1994 and 
April 1995.
The field of view of the $5 \times 5$ array was $200'' \times 200''$
($40'' \times 40''$ per pixel) and the beam size was $55''$ 
($\approx 8.3\times10^{-8}$~sr; FWHM).
We placed the array at 7 positions in M51.
Four positions were placed $160''$ apart in a square to cover the whole
galaxy.
To fully sample the central region, two positions were shifted diagonally
in north-east and south-west directions by 1.5 pixels with respect to the 
center and one position was placed at the center.
The center position ((0,0) position) is at RA = 
$13^{\rm h}27^{\rm m}46.4^{\rm s}$, Dec = $47^{\circ}27'13''$ (1950).
The total observed field of view was $360'' \times 360''$.
The position accuracy is estimated to be about $10''$.

Except for the position at the center all scans were carried out in 
``stare'' mode (fixed plate separation of the resolution determining
Fabry-Perot interferometer).
The spectral resolution in this mode was 120~km~s$^{-1}$ (FWHM).
To take the dispersion of the velocity shifts in M51 into account (more than
100~km~s$^{-1}$) we readjusted the plate separation of the Fabry-Perot 
interferometers for each individual array position.
However, the spectral range of some of the edge pixel which cover outer
parts of M51 do not cover the entire expected velocity range of the [C~{\small II}]
line.
Those pixels were not used for the data analyses.
For the center position we scanned the Fabry-Perot interferometer over
300~km~s$^{-1}$.
The spectral resolution at this position was 114~km~s$^{-1}$ (FWHM).
We used the H$_{2}$S line at 158.0148~$\mu$m to calibrate the transmission 
wavelength of the Fabry-Perot interferometers.

Internal blackbodies were used to flatfield the detector array.
We observed the planets Jupiter (June 1994) and Mars (April 1995) for 
absolute calibration of our data.
For the brightness temperatures of Jupiter at 158~$\mu$m we used
$T = 128$~K (Hildebrand et al.\ 1985).
We determined the brightness temperature of Mars to be 207~K at 158~$\mu$m
from the published values of Wright (1976) and Wright \& Odenwald (1980).
The mean diameter of Jupiter at the time of the observation was about 
$42''$ and the mean diameter of Mars was about $10''$.
The accuracy of the absolute calibration is about 30\%.

\begin{figure*}
\centering
\includegraphics[scale=0.60]{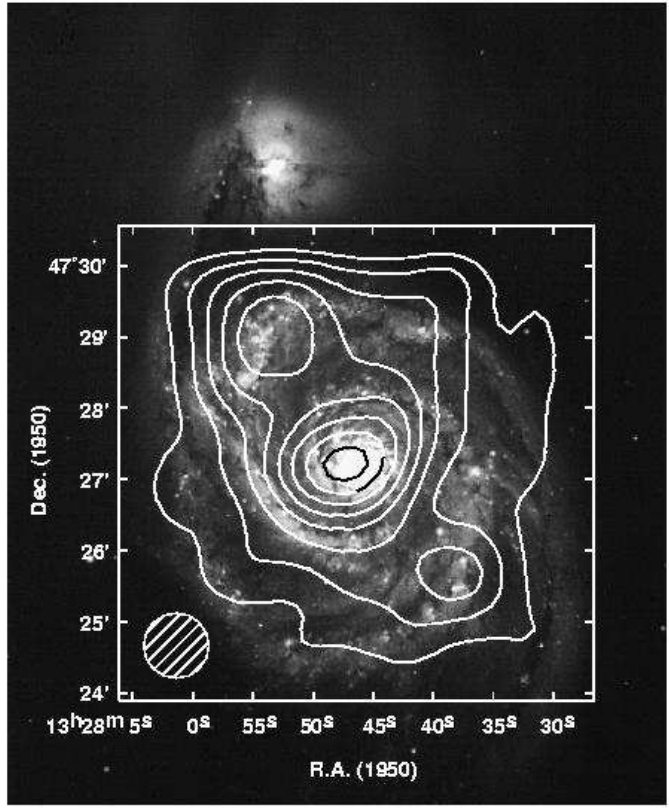}
\caption{
  Contour map of the [C~II] line superimposed on an optical image
  of M51.
  The peak intensity is  $I_{\rm [C II]} = 1.3 \times
  10^{-4}$ erg~s$^{-1}$cm$^{-2}$sr$^{-1}$ and the contour lines are in steps
  of $1.5 \times 10^{-5}$ erg~s$^{-1}$cm$^{-2}$sr$^{-1}$ starting at
  $1.5 \times 10^{-5}$ erg~s$^{-1}$cm$^{-2}$sr$^{-1}$.
  The hashed circle represents the beam size ($55''$ FWHM) of the [C~II]
  observation.
}
\label{fig1}
\end{figure*}

The final map of the integrated [C~{\small II}] line intensity 
(Figure~\ref{fig1}) was made using a maximum entropy image restoration
program.
This program was written especially for the FIFI data and follows the 
algorithm as outlined in Skilling \& Bryan (1984).
It takes the data from all positions observed  with FIFI and reconstructs
an image on a finer grid using the maximum entropy method and constrains
the data points by a chi-square fit.
The result is a deconvolved image which is then convolved with a $55''$
(FWHM) Gaussian beam.
An additional large error term was added to pixels which do not cover the
full velocity range within their beam.
This resulted in a smoothing effect due to a strong influence from
neighboring pixels which do cover the full veloctiy range.
The contour lines in the final map are smoother then in the raw map and
all features can be traced back to the observed data (no artificial
features were introduced by the program).
On the other hand all obvious features in the observed data appear in the
final map.

\section{Results}

\subsection{Morphology}

The contour map of the integrated [C~{\small II}] line intensity 
superimposed on an optical image of M51 (Figure~\ref{fig1}) shows 
that the [C~{\small II}] emission is distributed over the whole galaxy,
covering a total solid angle of $2.85\times10^{-6}$~sr and peaking at the 
nucleus.
The integrated [C~{\small II}] intensity at this peak is
$I_{\rm [C II]} = (1.31 \pm 0.15) \times 10^{-4}$ 
erg~s$^{-1}$cm$^{-2}$sr$^{-1}$.
This is in very good agreement with an earlier measurement of the 
integrated [C~{\small II}] intensity by Stacey et al.\ (1991)
who found $I_{\rm [C II]} = (1.4 \pm 0.3) \times 10^{-4}$ 
erg~s$^{-1}$cm$^{-2}$sr$^{-1}$ (Stacey 2001; see also Crawford et al.\
1985).
A comparison of the FIFI [C~{\small II}] results with our ISO
[C~{\small II}] data of M51 will be presented in a separate paper
(Nikola et al.\ 2002).
Preliminary reduction of our [C~{\small II}] ISO measurement of
M51 shows a lower intensity compared to the FIFI intensity, but roughly
consistent with the ratio of the beam sizes.

There is also a peak in the north-east (NE peak) of
M51 (at RA = $13^{\rm h}27^{\rm m}53.3^{\rm s}$, 
Dec = $47^{\circ}29'03''$ (1950))
coinciding with an enhancement in the optical emission in the
spiral arm.
The integrated [C~{\small II}] intensity at the NE peak is
$I_{\rm [C II]} = 9.2 \times 10^{-5}$ erg~s$^{-1}$cm$^{-2}$sr$^{-1}$ or 
about 70\% of the nuclear value.
A third peak in the [C~{\small II}] emission in the south-west (SW peak)
of M51 (at RA = $13^{\rm h}27^{\rm m}38.5^{\rm s}$, 
Dec = $47^{\circ}25'33''$ (1950))
coincides with a position of a spiral arm where enhanced
optical emission can be seen.
Here the integrated [C~{\small II}] intensity is
$I_{\rm [C II]} = 5.6 \times 10^{-5}$ erg~s$^{-1}$cm$^{-2}$sr$^{-1}$.

The NE and SW peaks are roughly symmetrically aligned with 
respect to the nucleus, and fall on local brightness peaks in the 
optical spiral arms.
Other than this, there is only a weak indication for spiral structure in 
the [C~{\small II}] map.
There is an extension in the [C~{\small II}] emission that follows the
northern spiral arm past the NE [C~{\small II}] peak, but the NE and SW
peaks appear to be quite localized, essentially point sources to our 
$55''$ beam. 

\subsection{Comparison with FIR Continuum}

\begin{figure*}
\centering
\includegraphics[scale=0.65]{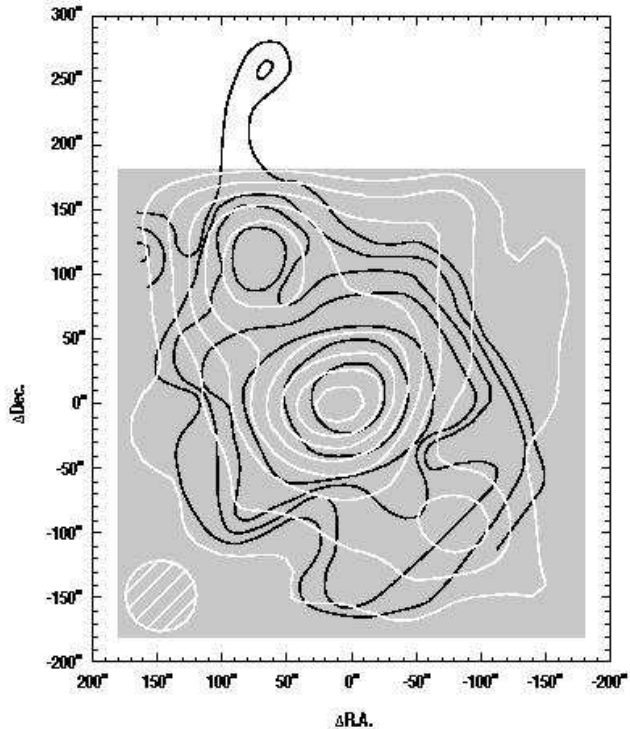}
\caption{
  Contour map of the $[$C~II$]$ emission (white) superimposed on the
  contour map of the FIR continuum emission (black) at 170~$\mu$m 
  (Smith 1982).
  The gray area outlines the observed field in the $[$C~II$]$ line.
  The contour lines of the FIR continuum are 9, 13, 18, 25, 35, 50, 
  and 70\% of the FIR peak ($8\times10^{-18}$ 
  W~m$^{-2}$Hz$^{-1}$sr$^{-1}$).
  The (0,0) position corresponds to 
  RA = $13^{\rm h}27^{\rm m}46.4^{\rm s}$, 
  Dec = $47^{\circ}27'13''$ (1950).
  The white hashed circle represents the beam size ($55''$ FWHM) of the 
  [C~II] observation.
  The spatial resolution of the FIR continuum map is $49''$.
}
\label{fig2}
\end{figure*}

A comparison of the distribution of the [C~{\small II}] emission with the
distribution of the FIR emission shows very good agreement.
Figure~\ref{fig2} shows the contour map of the integrated 
[C~{\small II}] line intensity superimposed on a contour map of the 
170~$\mu$m FIR continuum (Smith 1982) at a spatial 
resolution ($49''$) similar to that of the [C~{\small II}] map.
The location of the main peak and the NE peak in both distributions match 
each other very well.
There is no peak towards the SW, but there is a ridge from the nuclear
region that touches our SW [C~{\small II}] peak.
It is likely that these are the same structures, but not coincident due to
modest signal-to-noise ratios in both maps.

The integrated [C~{\small II}] luminosity of M51 is 
$L_{\rm [CII]} = 3 \times 10^{8} L_{\odot}$ or 1\% of the 
FIR luminosity as measured by the IRAS satellite
($3 \times 10^{10} L_{\odot}$, Rice et al.\ 1988, scaled to 9.6~Mpc).
This is at the high end of the $L_{\rm [CII]}$/$L_{\rm FIR}$ ratio observed
in galactic nuclei(Crawford et al.\ 1985; Stacey et al.\ 1991), but
consistent with the reported ratios for late type
galaxies (NGC~6946, Madden et al.\ 1993; NGC~4038/4039, Nikola
et al.\ 1998).
We use the Smith (1982) map to calculate the 
$L_{\rm [CII]}$/$L_{\rm FIR}$ ratio within our beam.
To do this, we correct the Smith (1982) luminosity estimates
(that include only the 80~$\mu$m and 200~$\mu$m contiunuum) to include
the shorter wavelength FIR continuum as observed by IRAS.
This correction amounts to a factor of 1.5  increase in the Smith (1982)
$L_{\rm IR}$ map.
The ratios range form 0.6\% at the nucleus to 1.1\% and 1.4\% respectively
at the NE and SW peaks (Table\ref{tab:c+fir}).

\begin{table*}
%\begin{center}
\centering
 \caption{Observed [C II] and FIR luminosity}
 \label{tab:c+fir}
\begin{tabular}{l|cccc}
 \hline\hline
 & Main [C~{\small II}] Peak & NE Peak & SW Peak & Total M51$^{\rm a)}$ \\
\hline
 $L_{\rm [CII]}$ [$L_{\odot}$] & $3.2 \times 10^{7}$ & $2.2 \times 10^{7}$ 
 & $1.4 \times 10^{7}$ & $3.0 \times 10^{8}$ \\
 $L_{\rm FIR}$$^{b)}$ [$L_{\odot}$] & $5.4 \times 10^{9}$ &
 $1.9 \times 10^{9}$ & $1.0 \times 10^{9}$ & $3.0 \times 10^{10}$ \\
 $L_{\rm [CII]}$/$L_{\rm FIR}$ & 0.6\% & 1.1\% & 1.4\% & 1.0\% \\
\hline
\end{tabular}
%\end{center}
\hspace*{1cm}\parbox{17cm}{\footnotesize
 $^{\rm a)}$ Averaged over the whole galaxy M51 
 ($\Omega \approx 2.85\times10^{-6}$~sr). \\
 $^{\rm b)}$ From Smith (1982) (corrected to include flux $<80~\mu$m using
  IRAS numbers)}
\end{table*}

\subsection{Comparison with CO (1$\to$0) Observations}

\begin{figure*}
\centering
\vspace{1cm}
\includegraphics[scale=0.65]{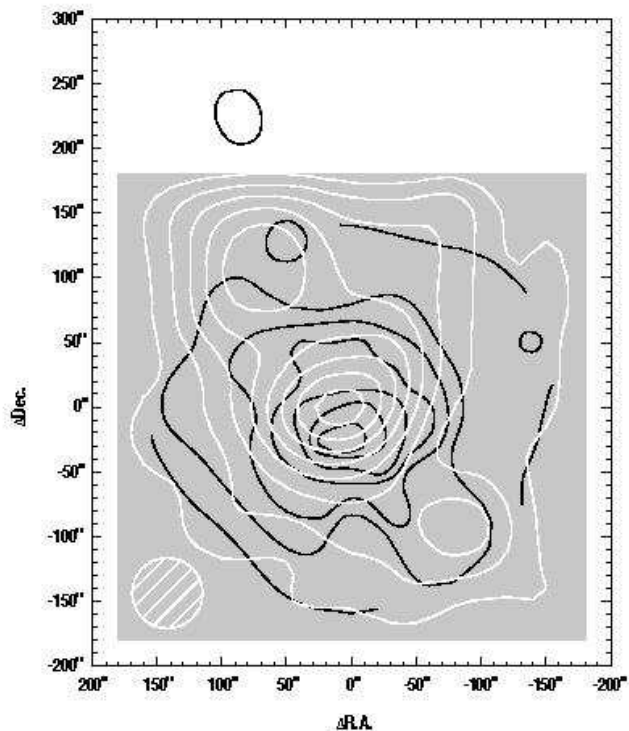}
\caption{
  Contour map of the the $[$C~II$]$ emission (white) superimposed
  on a contour map of the $^{12}$CO(1$\to$0) emission (black) derived
  from CO data from Lord \& Young (1990).
  The contours of the CO line are 5, 10, 15, 20, 30, 40, 50 K~km~s$^{-1}$ 
  (corrected for beam efficiency).
  The (0,0) position corresponds to 
  RA = $13^{\rm h}27^{\rm m}46.4^{\rm s}$, 
  Dec = $47^{\circ}27'13''$ (1950).
  The hashed circle represents the beam size ($55''$ FWHM) of the [C~II]
  observation.
}
\label{fig3}
\end{figure*}

As shown by Crawford et al.\ (1985) and Stacey et al.\ (1991)
the combination of the integrated [C~{\small II}] intensity with the
integrated intensity of the $^{12}$CO~(1$\to$0) line is a powerful
diagnostic tool for measuring the star formation activity.
To do this analysis we derived a CO~(1$\to$0) contour map from the CO data
of Lord \& Young (1990) because their observations have been carried out 
with a spatial resolution ($45''$) similar to our [C~{\small II}] 
observation.
However, their selection of positions does not cover the whole galaxy and we
interpolated over the narrow gaps in their observation.
Unfortunately their coverage shows a gap at the position of the NE peak.
Figure~\ref{fig3} shows the superposition of the 
[C~{\small II}] contour map on top of the derived contour map of the 
CO~(1$\to$0) line.
The peak integrated intensity of the CO emission is 
$I_{\rm CO} = 8.4 \times 10^{-8}$ erg~s$^{-1}$cm$^{-2}$sr$^{-1}$.
At the [C~{\small II}] peaks we derived integrated intensities of the
CO line of
$I_{\rm CO} = 6.6 \times 10^{-8}$ erg~s$^{-1}$cm$^{-2}$sr$^{-1}$
at the main [C~{\small II}] peak,
$I_{\rm CO} = 1.8 \times 10^{-8}$ erg~s$^{-1}$cm$^{-2}$sr$^{-1}$
at the NE peak, and
$I_{\rm CO} = 1.7 \times 10^{-8}$ erg~s$^{-1}$cm$^{-2}$sr$^{-1}$
at the SW peak.
The peak of the CO emission is not at the nominal nucleus of M51 and does 
not coincide with the main peak of the [C~{\small II}] emission.
Another observation of the CO~(1$\to$0) line by Scoville \& Young (1983) 
with the same spatial resolution but worse spatial coverage shows the peak of 
the CO emission in the center of the galaxy.
The CO~(1$\to$0) map of Garc\' \i a-Burillo, Gu\'elin, \& Cernicharo (1993),
with a spatial resolution of $33''$, shows a local minimum at the center and 
the CO peak shifted by about ($-20''$, $-20''$) relative to the center of M51.
This position is again in better agreement with the CO peak of Lord \& Young
(1990).
Apart from the slightly different positions of the CO peak and the
[C~{\small II}] main peak both distributions are reasonable similar.
The distribution of the CO emission is also elongated in north-east 
south-west direction and shows an extension toward the SW peak.
The overall agreement suggests much of the observed [C~{\small II}]
emission arises in PDRs on the surface of molecular clouds.

\subsection{Comparison with H$\alpha$ Emission}

\begin{figure*}
\centering
\includegraphics[scale=0.65]{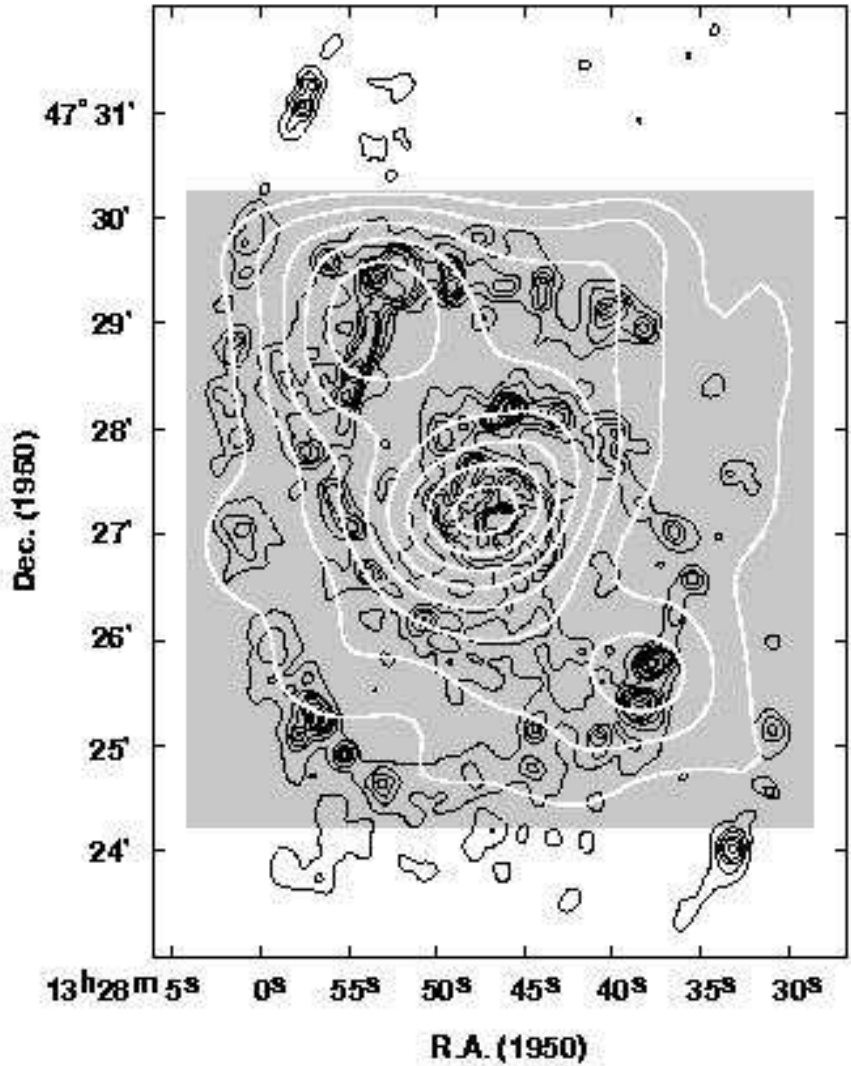}
\caption{
  Contour map of the $[$C~II$]$ emission (white) superimposed on a
  contour map of the H$\alpha$ emission (black) (van der Hulst et al.\
  1988).
  The spatial resolution of the H$\alpha$ map is $8''$ and the contours
  are 0.05, 0.25, 0.5, 0.75, 1.0, 1.5, 2.0, 3.0, 4.0 $\times 10^{-13}$ 
  erg~s$^{-1}$cm$^{-2}$.
}
\label{fig4}
\end{figure*}

Figure~\ref{fig4} shows the superposition of the [C~{\small II}]
contour map on the H$\alpha$ contour map of van der Hulst et al.\ 
(1988).
Despite the very different spatial resolution ($8''$ for the H$\alpha$ map)
the maps are similar in the sense that the
NE peak and the SW peak coincide with an accumulation of strong,
discrete H$\alpha$ knots in the spiral arms.
There are three moderately strong H$\alpha$ peaks located in the spiral arm 
in the south-east where we only detect weak [C~{\small II}] emission.
This might be the result of incomplete coverage of the full velocity 
dispersion at the edge of our array as described in section~\ref{observation}.
Pixels showing this effect have not been taken into account for the data 
reduction.
These ``missing'' pixels created an incomplete image plane near the edge of
our field of view and can therefore easily cause the wavy contour line.
Apart from this, the superposition of the [C~{\small II}] map on the
H$\alpha$ map shows nicely that the [C~{\small II}] emission peaks coincide 
with strong H~{\small II} regions and that the [C~{\small II}] emission 
therefore traces mainly PDRs in star formation regions.
However, weak [C~{\small II}] emission is also seen in regions where there
is almost no H$\alpha$ emission, e.g.\ in the west-north-west region of M51.

\section{Origin of the [C~{\small II}] emission in M51}

\subsection{Cold and warm neutral medium}

\begin{figure*}
\centering
\includegraphics[scale=0.65]{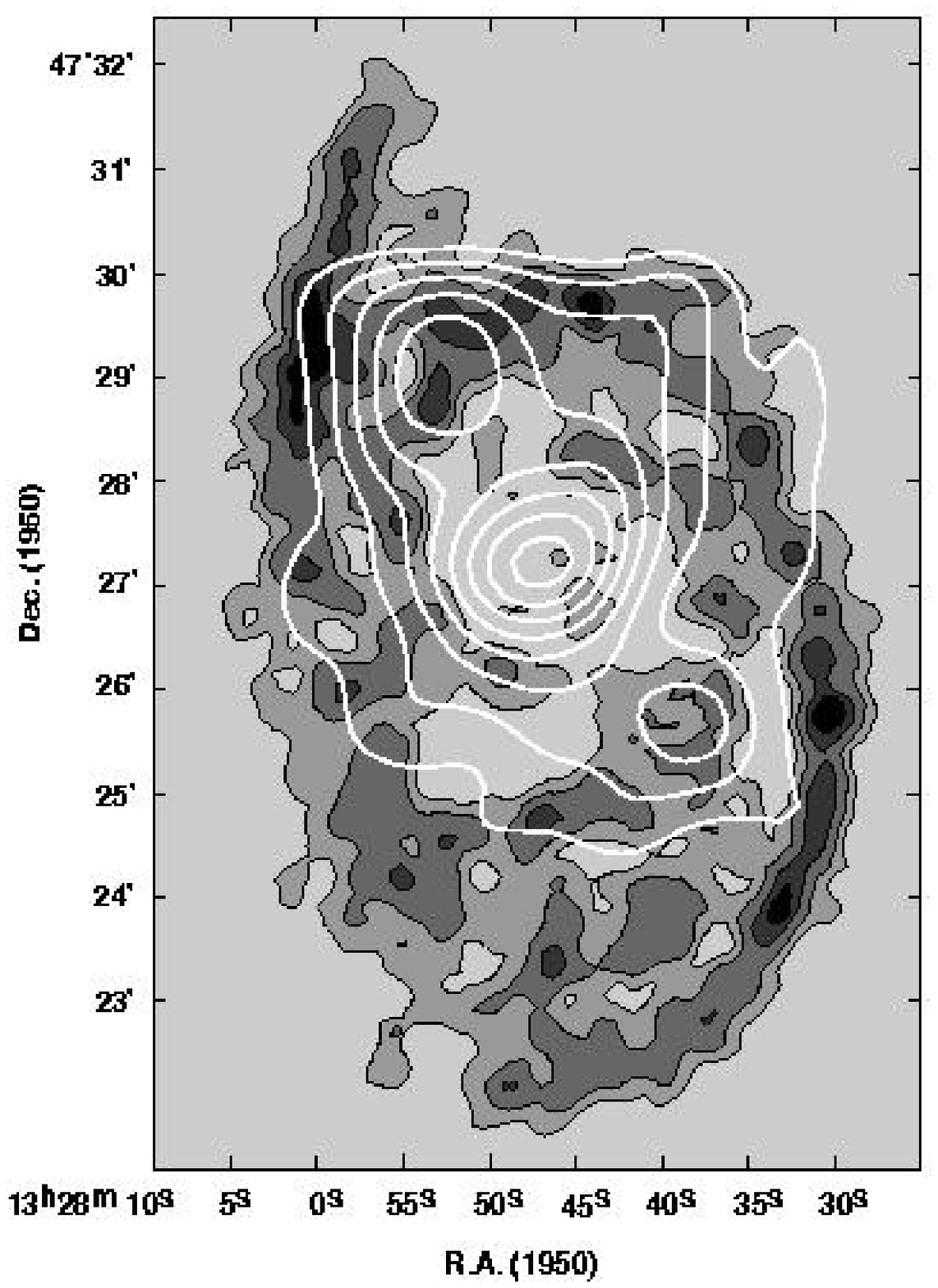}
\caption{
  Contour map of the $[$C~II$]$ emission (white) superimposed on a
  contour map of the H~I column density (gray scale) (Tilanus \& Allen
  1991).
  The spatial resolution of the H~I map is $20''$ and the contours
  are 3, 6, 10.5, 14.9 $\times 10^{20}$ atoms~cm$^{-2}$.
  Darker colors correspond to higher column densities.
}
\label{fig5}
\end{figure*}

Comparison of the H~{\small I} column density maps of Weliachew \& 
Gottesman (1973) and Tilanus \& Allen (1991) (Figure~\ref{fig5}) with our 
[C~{\small II}] map shows that the distributions do not agree.
The H~{\small I} column density has a minimum at the center of M51 and 
shows a ridge of higher column density around the nucleus.
On top of the H~{\small I} column density ridge are several local column
density maxima.
The NE or SW peaks lie on this H~{\small I} column density ridge but do not
coincide with the local column density maxima.
Although local H~{\small I} column density maxima are in the vicinity of 
the NE and
SW peak the overall morphology of the [C~{\small II}] distribution and the
H~{\small I} column denisty distributions do not agree.
It therefore seems unlikely that much of the peak [C~{\small II}] emission 
arises from diffuse neutral medium.

Here we calculate the expected integrated [C~{\small II}] intensity from 
the cold and warm atomic medium.
For the calculation we assume that C$^{+}$ behaves like a two-level system
and that the [C~{\small II}] radiation is optically thin.
This gives:
\begin{displaymath}
 \begin{array}{rcl}
  I_{\rm [C II]} & = & \frac{h \nu A}{4 \pi} \left[
  \frac{\frac{g_{u}}{g_{l}} \exp(- \frac{91}{T})}{1 + \frac{g_{u}}{g_{l}} 
  \exp(- \frac{91}{T}) +  \frac{n_{\rm cr, H}}{n_{\rm H}}} + \right. \\
 & & \left. + \frac{\frac{g_{u}}{g_{l}} \exp(- \frac{91}{T})}{1 + 
     \frac{g_{u}}{g_{l}}
  \exp(- \frac{91}{T}) + \frac{n_{\rm cr, e}}{n_{\rm H} x_{\rm e}}} 
  \right] {X_{\rm C^{+}} N_{\rm H} \Phi_{\rm B}}
 \end{array}
\end{displaymath}
(see also Madden et al.\ 1993, 1997)
where $h$ is the Planck constant, $\nu$ is the frequency of the transition, 
$A$ is the Einstein coefficient for spontaneous emission, $2.29 \times 
10^{-6}$ s$^{-1}$ (Nussbaumer \& Storey 1981), and
$g_{u}/g_{l} = 2$ is the ratio of the statistical weights in the 
upper and lower level.
The first term in the equation above is due to excitation by hydrogen 
impacts 
while the second term is due to excitation by electron impacts.
The critical density for collisions with atomic hydrogen, $n_{\rm cr, H}$,
is deduced from the cooling function 
for collisional excitation of [C~{\small II}] by atomic hydrogen 
(Launay \& Roueff 1977), and that for collisions with electrons, 
$n_{\rm cr, e}$, 
is deduced from collision strengths for [C~{\small II}] collisions with 
electrons (Blum \& Pradhan 1992), and fitted as functions of temperature.
We further assume a gas phase abundance of carbon relative to hydrogen of 
$X_{\rm C^{+}} = 3 \times 10^{-4}$, that
all the carbon is ionized, a beam filling factor of $\Phi_{\rm B} = 1$, 
and that the same range of conditions for the cold 
neutral medium (CNM temperature $T = 50 - 100$~K, density $n_{\rm H} = 
50 - 200$~cm$^{-3}$ 
and ionization fraction $x_{\rm e} = 3 \times 10^{-4})$ and the warm neutral
medium (WNM temperature $T = 4000 - 8000$~K, density $n_{\rm H} = 
0.5 - 5$~cm$^{-3}$ and 
ionization fraction $x_{\rm e} = 3 \times 10^{-2}$) (Kulkarni \& Heiles 
1987; Kulkarni \& Heiles 1988) of the Milky Way apply for M51.
We derived the H~{\small I} column density for our calculation from the
gas surface density values of Lord \& Young (1990) who
used H~{\small I} data of Rots et al.\ (1990) smoothed to a resolution
of $45''$ for their estimate.
The results are given in Table~\ref{tab:cnm_wnm}.

\begin{table*}
%\begin{center}
\centering
 \caption{Observed and estimated integrated [C II] intensity from the CNM, 
 the WNM, and the ELDWIM}
 \label{tab:cnm_wnm}
 \vspace{0.5cm}
{\small
\begin{tabular}{rl|cccc}
 \hline\hline
 & & \multicolumn{4}{c}{$I_{\rm [CII]}$ in 
 [erg~s$^{-1}$cm$^{-2}$sr$^{-1}$]} \\
 \multicolumn{2}{c|}{} & Main [C~{\small II}] Peak & NE Peak & SW 
 Peak & Total M51$^{\rm a)}$ \\
\hline
 \multicolumn{2}{c|}{Observed $I_{\rm [CII]}$} & $1.31 \times 10^{-4}$ & 
 $9.2 \times 10^{-5}$ & $5.6 \times 10^{-5}$ & $3.7 \times 10^{-5}$ \\
 & & \multicolumn{4}{c}{} \\
 \multicolumn{2}{c|}{$N_{\rm H}$~[cm$^{-2}$]} & $9.6 \times 10^{20}$ & 
 $1.3 \times 10^{21}$ & $1.2 \times 10^{21}$ & $1.1 \times 10^{21}$ \\
 & & \multicolumn{4}{c}{} \\
 & $I_{\rm [CII]min}$$^{\rm b)}$ & $1.0 \times 10^{-5}$ (8\%)$^{\rm e)}$ &
 $1.4 \times 10^{-5}$ (15\%) & $1.3 \times 10^{-5}$ 
 (23\%) & $1.2 \times 10^{-5}$ (32\%) \\
 CNM & $I_{\rm [CII]std}$$^{\rm b)}$ & $1.6 \times 10^{-5}$ 
 (12\%) & $2.2 \times 10^{-5}$ (24\%) & 
 $2.0 \times 10^{-5}$ (36\%) & 
 $1.8 \times 10^{-5}$ (49\%) \\
 & $I_{\rm [CII]max}$$^{\rm b)}$ & $1.6 \times 10^{-5}$ (12\%) 
 & $2.2 \times 10^{-5}$ (24\%) & $2.0 \times 10^{-5}$ 
 (36\%) & $1.8 \times 10^{-5}$ (49\%) \\
 & & \multicolumn{4}{c}{} \\
 & $I_{\rm [CII]min}$$^{\rm c)}$ & $8.4 \times 10^{-7}$ (0.6\%) 
 & $1.1 \times 10^{-6}$ (1\%) & $1.1 \times 10^{-6}$ 
 (2\%) & $9.7 \times 10^{-7}$ (3\%) \\
 WNM & $I_{\rm [CII]std}$$^{\rm c)}$ & $1.7 \times 10^{-6}$ 
 (1\%) & $2.3 \times 10^{-6}$ (2.5\%) & 
 $2.1 \times 10^{-6}$ (4\%) & $2.0 \times 10^{-6}$ (5\%) \\
 & $I_{\rm [CII]max}$$^{\rm c)}$ & $8.7 \times 10^{-6}$ (7\%) & 
 $1.2 \times 10^{-5}$ (13\%) & $1.1 \times 10^{-5}$ 
 (20\%) & $9.9 \times 10^{-6}$ (27\%) \\
& & \multicolumn{4}{c}{} \\
 \multicolumn{2}{c|}{$EM$~[pc~cm$^{-6}$]} & 1387 & 412 & 653 & 91 \\
& & \multicolumn{4}{c}{} \\
 & $I_{\rm [CII]min}$$^{\rm d)}$ & $3.9 \times 10^{-5}$ (30\%) & 
 $6.6 \times 10^{-6}$ (7.2\%) & $9.4 \times 10^{-6}$ 
 (16.8\%) & $7.8 \times 10^{-6}$ (21\%) \\ 
\raisebox{1.5ex}[-1.5ex]{ELDWIM} & $I_{\rm [CII]max}$$^{\rm d)}$ & 
 ($> 100$\%) & $3.9 \times 10^{-5}$ (42\%) & 
 ($> 100$\%) & ($> 100$\%) \\
 \hline
\end{tabular}
}
%\end{center}
\hspace*{0.1cm}\parbox{17cm}{\footnotesize
 $^{\rm a)}$ Averaged over the whole galaxy M51 
 ($\Omega \approx 2.85\times10^{-6}$~sr). \\
 $^{\rm b)}$ Temperatures and densities used for CNM: \\
 \hspace*{1em}min.: $n_{\rm H} = 50$~cm$^{-3}$, $T = 100$~K;
 max.: $n_{\rm H} = 200$~cm$^{-3}$, $T = 50$~K;
standard: $n_{\rm H} = 100$~cm$^{-3}$, $T = 80$~K; \\
 $^{\rm c)}$ Temperatures and densities used for WNM: \\
 \hspace*{1em}min.: $n_{\rm H} = 0.5$~cm$^{-3}$, $T = 8000$~K;
 max.: $n_{\rm H} = 5$~cm$^{-3}$, $T = 4000$~K;
standard: $n_{\rm H} = 1$~cm$^{-3}$, $T = 6000$~K; \\
 $^{\rm d)}$ see text for minimum and maximum contribution \\
 $^{\rm e)}$ Fractions of observed integrated [C~{\small II}] intensity are
 noted in parenthesis.}
\end{table*}

Clearly, emission from the WNM is not important for most of the regions 
mapped in [C~{\small II}].
Even the maximum values for the expected integrated [C~{\small II}]
intensity from the WNM are below the [C~{\small II}] detection limit.

A non-negligible contribution to the [C~{\small II}] emission from the WNM 
is only possible in the north-west where extended [C~{\small II}] emission 
at the detection limit is visible.
At this position the H~{\small I} column density derived from the data of
Lord \& Young (1990) reaches its maximum value of
$N_{\rm H} \approx 2.5 \times 10^{21}$~cm$^{-2}$. 
The estimated contribution of the WNM in this region goes from a minimum
of $2.2 \times 10^{-6}$~erg~s$^{-1}$cm$^{-2}$sr$^{-1}$ or about 7\% of the 
integrated [C~{\small II}] intensity ($\lesssim 
3 \times 10^{-5}$~erg~s$^{-1}$cm$^{-2}$sr$^{-1}$) up to a maximum of
$2.3 \times 10^{-5}$~erg~s$^{-1}$cm$^{-2}$sr$^{-1}$ or 77\% over the range
of the physical conditions of the WNM.
In this region, in the north-west, the
contribution of the CNM to the integrated [C~{\small II}] intensity
can easily account for the total observed intensity in the north-west
region.
Even the minimum contribution from the CNM 
($2.9 \times 10^{-5}$~erg~s$^{-1}$cm$^{-2}$sr$^{-1}$) can account for all
of the observed [C~{\small II}] emission.
At standard conditions ($n_{\rm H} \sim 100$~cm$^{-3}$, $T \sim 80$~K) 
the contribution from CNM to the [C~{\small II}] emission is
$4.6 \times 10^{-5}$~erg~s$^{-1}$cm$^{-2}$sr$^{-1}$ which is more than
1.5 times the observed [C~{\small II}] emission.

At the [C~{\small II}] peak positions the estimated integrated 
[C~{\small II}] intensity arising from the CNM can also be relatively 
high.
The estimated fraction of [C~{\small II}] emission coming from the CNM
at the NE and SW peak is between $\sim 15$\% and $\sim 36$\% (see 
Table~\ref{tab:cnm_wnm}).
Although the CNM contribution is smaller at the position of the main
[C~{\small II}] peak it is still not negligible
(8 -- 12\%; see Table~\ref{tab:cnm_wnm}).

Since we assumed that all carbon in the CNM and WNM is in the form C$^{+}$
and that all of the observed H~{\small I} column density is associated with
the [C~{\small II}] emission, the estimated values are upper limits.
Therefore a contribution to the [C~{\small II}] emission form the CNM is
likely smaller than that called out above, so that at the nucleus , NE and
SW peaks, the contribution may be small ($<$10 -- 20\%).
However, in the north-west region of M51 the CNM could be
the dominant source of the [C~{\small II}] emission.

\subsection{Ionized medium}

The estimate of the contribution of the ionized medium to the
[C~{\small II}] emission is very controversial.
We will discuss four different approaches here.
The easiest approaches are similar to the one we used to estimate the 
contribution of the CNM and WNM.
This leads to the following formula
\begin{displaymath}
 \begin{array}{rcl}
  I_{\rm [C II]} & = & \frac{h \nu A}{4 \pi n_{\rm cr, e}} \left[
  \frac{\frac{g_{u}}{g_{l}} \exp(- \frac{91}{T})}{1 + \left( 1 + 
  \frac{g_{u}}{g_{l}} \exp(- \frac{91}{T}) \right)
  \frac{n_{\rm e}}{n_{\rm cr, e}}} \right] \times \\
  & & \times X_{\rm C^{+}} \Phi_{\rm B}
  \int_{0}^{s'} n_{\rm H} n_{\rm e} {\rm d}s 
 \end{array}
\end{displaymath}
where $n_{\rm cr, e} \approx 49$ cm$^{-3}$ is the 
critical density of C$^{+}$ for collisions with electrons at 8000~K 
(Blum \& Pradhan 1992), and 
$\int_{0}^{s'} n_{\rm H} n_{\rm e} {\rm d}s$ is the emission measure (EM).

Klein, Wielebinski, \& Beck (1984) and van der Hulst et al.\ (1988)
determined the flux density of the thermal radio continuum from single-dish 
measurements with a beam size of $76''$ and from interferometric 
measurements with a spatial resolution of $8''$, respectively.
The thermal flux density derived from the interferometric measurement 
is about 34\% of that derived from the single-dish observation and
originates mainly from prominent giant H~{\small II} regions 
with sizes of about 500~pc and electron densities of about $n_{\rm e} 
\approx 1 - 2$~cm$^{-3}$.
These H~{\small II} clouds are distributed within the nuclear region of 
M51 and along the spiral arms.
In contrast, the distribution of the thermal 2~cm radio continuum derived 
from the single-dish observation does not show any spiral signature.

Due to the big beam size of FIFI the main contributer of the ionized medium 
within the beam is most likely the extended low density warm ionized 
medium (ELDWIM).
In our first approach we
therefore estimate a possible [C~{\small II}] emission from ionized gas
using the single-dish measurements of the radio continuum of Klein, 
Wielebinski, \& Beck (1984) assuming that all the thermal radio continuum
arises from the ELDWIM.
In this case the intensity of the thermal radio continuum is the same
within the FIFI beam as within the single-dish beam.
From the emission measure, derived from the thermal radio continuum, and
assuming the low density limit ($n_{\rm e} \ll n_{\rm cr}$) and a beam 
filling factor of unity we estimate the contribution of the ELDWIM to the 
observed [C~{\small II}] intensity from the above formula.
The flux density of the thermal radio continuum at the NE  and SW peak are 
taken from the 2~cm thermal continuum contour map of Klein, Wielebinski, \& 
Beck (1984; their Figure~7).
The intensity of the thermal radio continuum at 2~cm ($T_{\rm 2th}$) at 
the NE peak is $T_{\rm 2th} \approx 2.5$~mK.
From this we get an emission measure of $EM \approx 215$~pc~cm$^{-6}$
(see Spitzer 1978 for the calculation) resulting in an expected integrated 
[C~{\small II}] intensity from ELDWIM at the NE peak of 
$I_{\rm [CII]} \approx 1.8 \times 10^{-5}$ erg~s$^{-1}$cm$^{-2}$sr$^{-1}$.
This is about 20\% of the observed integrated [C~{\small II}] intensity.
At the SW peak we get an intensity of the thermal radio continuum at 2~cm
of $T_{\rm 2th} \approx 4$~mK from the contour map in Klein, Wielebinski,
\& Beck (1984).
This leads to an emission measure of $EM \approx 344$~pc~cm$^{-6}$ and an 
integrated [C~{\small II}] intensity of 
$I_{\rm [CII]} \approx 2.9 \times 10^{-5}$ erg~s$^{-1}$cm$^{-2}$sr$^{-1}$
which is about 52\% of the observed integrated intensity.
Within an area of $60'' \times 70''$ at the center of M51, Klein, 
Wielebinski, \& Beck (1984) determined a thermal radio continuum at 2~cm of 
11~mJy.
Assuming again the same intensity within the FIFI beam as within this
area we calculate an emission measure of $EM \approx 1382$~pc~cm$^{-6}$ and
an expected integrated [C~{\small II}] intensity of 
$I_{\rm [CII]} \approx 1.18 \times 10^{-4}$ erg~s$^{-1}$cm$^{-2}$sr$^{-1}$.
This is about 90\% of the observed integrated [C~{\small II}] intensity.
The flux density of the thermal 2~cm radio continuum of the whole galaxy
M51 is $S_{\rm 2th} \approx 55$~mJy (Klein, Wielebinsky, \& Beck 1984).
This flux density was measured within an elliptical area with half axis of
$5.75'$ and $4.93'$.
The corresponding emission measure is $EM \approx 91$~pc~cm$^{-6}$.
This results in an expected mean integrated [C~{\small II}] intensity from
H~{\small II} regions of 
$I_{\rm [CII]} \approx 7.8 \times 10^{-6}$ erg~s$^{-1}$cm$^{-2}$sr$^{-1}$
for the whole galaxy which is about 21\% of the observed mean integrated
[C~{\small II}] intensity.

In the nuclear region and the north-east region the interferometer map of 
the 6~cm radio continuum of van der Hulst et al.\ (1988) reveals numerous
strong sources.
Thus the assumption that all the thermal radio continuum radiation in these
regions arises from ELDWIM might be too simple.
In our second approach we therefore assume that 34\% of the thermal 
radio continuum arises from clumpy H~{\small II} regions as indicated
in the radio continuum map of van der Hulst et al.\ (1988).
Correcting for beam filling from the $76''$ beam to the $55''$ beam and 
assuming an electron density of $n_{\rm e} = 1.5$~cm$^{-3}$ for the clumpy
medium enhances the expected integrated [C~{\small II}] intensity
originating from the clumpy medium and ELDWIM by a factor of 1.26 compared 
to the first approach.

Except for the NE peak and the total galaxy both approaches result in
large fractions of the [C~{\small II}] emission attributed to ELDWIM 
reaching 100\% for the nucleus.
However, by assuming a beam filling factor of unity and the low density
limit these values are upper limits.

Another approach to estimate the [C~{\small II}] emission from ELDWIM is 
to compare the integrated [C~{\small II}] intensity
with integrated [N~{\small II}] intensity.
The [N~{\small II}] emission comes entirely from ionized medium and the
critical densities of the [N~{\small II}] emission lines are similar to the
critical density of the [C~{\small II}].
Treating the N$^{+}$ ion as a pure three level system and 
assuming an abundance ratio for $X$(C$^{+}$)/$X$(N$^{+}$) it is possible
to estimate the contribution of the ELDWIM to the [C~{\small II}] emission
from [N~{\small II}] measurements.
In the low density limit the expected ratio of the integrated 
[C~{\small II}] intensity to the integrated [N~{\small II}] 122$\mu$m 
intensity from ELDWIM is
\begin{eqnarray}
\label{ic/in}
 \frac{I_{\rm [CII]}}{I_{\rm [NII],122}} & = & \frac{ 121.898 }{ 157.741 } 
 \times \frac{ A_{158} }{ A_{122} } \times \frac{ n_{\rm cr, e}^{122} }{ 
 n_{\rm cr, e}^{158} } \times \nonumber \\
& & \times \frac{ 2 \exp \left( - \frac{ 91 }{ 10^{4} } \right) 
 }{ \frac{ \gamma_{02} }{ \gamma_{21} + \gamma_{20} } } \times
 \frac{ X_{\rm C^{+}} }{ X_{\rm N^{+}} } \nonumber \\
& = & \frac{ 121.898 }{ 157.741 } \times 
 \frac{ 2.29 \times 10^{-6} }{ 7.46 \times 10^{-6} } \times
 \frac{ 310 }{ 49 } \times \nonumber \\
& &  \times \frac{ 
  1.98 
 }{ 
  \frac{ 
   2.3 \times 10^{-8} 
  }{ 
   1.9 \times 10^{-8} + 4.7 \times 10^{-9} 
  }
 } \ 
 \frac{ 
  3 \times 10^{-4} 
 }{ 
  1 \times 10^{-4} 
 } \nonumber \\
& \approx & 9
\end{eqnarray}
where $A_{158}$ and $A_{122}$ are the emission coefficients for
spontaneous emission of the [C~{\small II}] line and the 122~$\mu$m 
[N~{\small II}] line, respectively (Nussbaumer \& Storey 1981, Nussbaumer 
\& Rusca 1979), the
$\gamma_{\rm ij}$'s are the collision rate coefficients derived from the
effective collision strengths given by Lennon \& Burke (1994), and
$n_{\rm cr, e}^{122}$ and $n_{\rm cr, e}^{158}$ are the critical densities 
for collisions with electrons at a temperature of $10^{4}$~K
for the [C~{\small II}] line and the 122~$\mu$m 
[N~{\small II}] line, respectively.
The contribution of the ELDWIM to the observed [C~{\small II}] emission
is then given by $(9 \times I_{\rm [NII], 122}^{\rm observed}) / 
I_{\rm [CII]}^{\rm observed}$.
For the comparison it is preferable to have the [N~{\small II}] and 
[C~{\small II}] intensities obtained with the same instrument.
Using our ISO [N~{\small II}] 122~$\mu$m and ISO [C~{\small II}] data
(Nikola et al.\ 2002)
we obtain a contribution of the ELDWIM to the observed ISO [C~{\small II}] 
emission of $(9 \times 1.9 \times 10^{-6}) / 3.9 \times 10^{-5} \approx 
44\%$ for the NE peak, $(9 \times 1.4 \times 10^{-5}) / 7.2 \times 10^{-5} 
\approx 175\%$ for the nucleus, and $(9 \times 2.7 \times 10^{-6}) / 
2.5 \times 10^{-5} \approx 97\%$ for the SW peak, where the 
intensities are in units of ${\rm [erg~s^{-1}~cm^{-2}~sr^{-1}]}$.
These numbers are higher than the values derived above in our first two
approaches.
Note that the calibration of our ISO data is still uncertain.
Therefore the individual ISO intensities should only be used with 
reservations.
However, the ratio of the ISO intensities should not be affected by the
calibration uncertainty.
The relative statistical errors of the ISO [N~{\small II}] intensities are
$\approx 30\%$ at the NE- and SW-peak and $\approx 5\%$ at the nucleus,
and the relative statistical errors of the ISO [C~{\small II}] intensities
are $\approx 10\%$ at the NE-peak, $\approx 4\%$ at the SW-peak, and
$\approx 2\%$ at the nucleus.

From recent COBE FIRAS observation Petuchowski \& Bennett (1993)
estimated the morphology of the ionized medium in the Milky Way.
They estimate that only a small fraction (17\%) of the [N~{\small II}] 
122$\mu$m emission actually arises from ELDWIM and that the majority
(82\%) of the [N~{\small II}] emission comes from externally ionized
cloud surfaces.
These externally ionized cloud surfaces have higher densities 
($n_{\rm H} = 150$~cm$^{-3}$) than the ELDWIM.
If we assume the same conditions in M51 as in the Milky Way then
the low density limit used in the above calculation might therefore not
be appropriate.
Assuming again a pure three level system for the N$^{+}$ ion and a pure
two level system for the C$^{+}$ ion the expected intensity ratio of the 
158~$\mu$m line to the 122~$\mu$m line in an ionized medium with a density 
of 150~cm$^{-3}$ and a temperature of $10^{4}$~K decreases to
$I_{158}/I_{122} \approx 2.8$.
If we divide the [N~{\small II}] emission into a part (17\%) arising from 
ELDWIM and a part (82\%) arising from more dense externally ionized
cloud surfaces and using our ISO data we obtain a much lower contribution 
of the ELDWIM to the observed [C~{\small II}] emission:
7.2\% for the NE peak, 30\% for the nucleus, and 16.8\% for the SW peak.
For reference Petuchowski \& Bennett (1993) 
estimate that the ELDWIM in the Milky Way contributes about 53\% to 
the [C~{\small II}] emission.
This fraction is higher than the fractions we derive using the estimated
[N~{\small II}] morphology from COBE for all our positions but similar
or smaller than the fraction we estimated for the SW peak and the nucleus
in the first three approaches.
For the NE peak this contribution would be higher than the values derived
above in all four approaches.

For the minimum ELDWIM contribution to the [C~{\small II}] emission we 
listed the estimated integrated [C~{\small II}] intensity derived from the 
COBE FIRAS [N~{\small II}] 122~$\mu$m morphology (Table~\ref{tab:cnm_wnm}).
For the  total galaxy M51 the only way to estimate the minimum value was
by our first approach.
All the maximum contribution from ELDWIM are higher than 100\%,
except for the NE peak.
We got the maximum contribution to the [C~{\small II}] emission at this 
position when we assumed that all the ISO [N~{\small II}] emission arises
from ELDWIM.

In the further analysis we will use the minimum [C~{\small II}] 
contribution from the ELDWIM.

\subsection{Photodissociation regions}

Strong [C~{\small II}] emission is commonly attributed largely to PDRs
associated with molecular cloud surfaces.
However, as shown in the previous section in the case of M51 components 
other than PDRs can account for a large fraction of the observed 
integrated [C~{\small II}] intensity.
But presently it is not possible to make firm statements about how
much of the [C~{\small II}] emission is really coming from other components 
of the interstellar medium other than PDRs.
Therefore we will follow four ways in analyzing the [C~{\small II}] 
emission from PDRs.
First we assume that all the observed integrated [C~{\small II}] 
intensity originates from PDRs (``case-1'').
Then we subtract the minimum contribution of all non-PDR contributions from 
the observed integrated [C~{\small II}] intensity (``case-2'').
As a third case we subtract the standard contributions of the CNM and the 
WNM and the minium contribution of the ELDWIM from the observed 
[C~{\small II}] intensity and attribute the remaining intensity to 
PDRs (``case-3'').
For the last case we subtract the maximum contribution from CNM, WNM, 
and ELDWIM from the observed integrated intensity (``case-4'').
In this case the NE peak is the only position where a PDR contribution to
the [C~{\small II}] emission is left over.

\subsubsection{Determination of density, FUV intensity, and beam filling
factor of PDRs}

To determine the density, the far-UV (FUV) intensity and the beam filling 
factor of the PDRs we compare the integrated [C~{\small II}] intensity with 
the integrated $^{12}$CO(1$\to$0) intensity from Lord \& Young (1990) and 
the intensity of the FIR continuum from Smith (1982) as corrected
using IRAS data and relate this to PDR model predictions (Wolfire et al.\
1989, Stacey et al.\ 1991).
We determine these values for the three points of interest, the main
[C~{\small II}] peak, the NE peak, and the SW peak.
In addition we determine the mean values for the entire galaxy M51.
The values used in this estimate are given in Table~\ref{tab:pdr1}.

\begin{table*}
%\begin{center}
\centering
 \caption{Paramaters used for the PDR model}
 \label{tab:pdr1}
 \vspace{0.2cm}
{\small 
\begin{tabular}{r|cccc}
 \hline\hline
 & Main $[$C~II$]$ peak & NE peak & SW peak & Total M51$^{\rm a)}$ \\ 
 \hline
 $I_{\rm CO}$$^{\rm b)c)}$ & $6.6\times10^{-8}$ & 
 $1.8\times10^{-8}$ & $1.7\times10^{-8}$ & $2.8\times10^{-8}$ \\
 $\chi_{\rm FIR}$$^{\rm d)}$ & 112 & 40 & 20 & 23 \\
 & \multicolumn{4}{c}{} \\
 \multicolumn{1}{c|}{} & \multicolumn{4}{c}{Observed} \\
 $I_{\rm [CII]}$$^{\rm b)}$ & $1.31\times10^{-4}$ &
 $9.2\times10^{-5}$ & $5.6\times10^{-5}$ & $3.7\times10^{-5}$ \\
 $I_{\rm [CII]}$/$I_{\rm CO}$ & 2000 & 5000 & 3300 & 1300 \\
 & \multicolumn{4}{c}{} \\
 \multicolumn{1}{c|}{} & \multicolumn{4}{c}{Minimum Contribution from
 non-PDRs subtracted} \\
 $I_{\rm [CII]}$$^{\rm b)}$ & $8.0 \times 10^{-5}$ &
 $7.0 \times 10^{-5}$ & $3.1 \times 10^{-5}$ & $1.6 \times 10^{-5}$ \\
 & \multicolumn{4}{c}{} \\
 \multicolumn{1}{c|}{} & \multicolumn{4}{c}{Standard Contribution from
 the CNM and the WNM} \\
 \multicolumn{1}{c|}{} & \multicolumn{4}{c}{and minimum of ELDWIM 
 subtracted} \\
 $I_{\rm [CII]}$$^{\rm b)}$ & $7.3 \times 10^{-5}$ &
 $6.0 \times 10^{-5}$ & $2.3 \times 10^{-5}$ & $8.5 \times 10^{-6}$ \\
 & \multicolumn{4}{c}{} \\
 \multicolumn{1}{c|}{} & \multicolumn{4}{c}{Maximum Contribution from
 non-PDRs subtracted} \\
 $I_{\rm [CII]}$$^{\rm b)}$ & 0 & $1.7 \times 10^{-5}$ & 0 & 0 \\
 \hline 
\end{tabular}
}
%\end{center}
\hspace*{2cm}\parbox{15cm}{\footnotesize
 $^{\rm a)}$ Averaged over the whole galaxy M51 
 ($\Omega \approx 2.85\times10^{-6}$~sr). \\
 $^{\rm b)}$ The integrated intensities are in units of
 [erg~s$^{-1}$cm$^{-2}$sr$^{-1}$]. \\
 $^{\rm c)}$ $I_{\rm CO} = 1.6 \times 10^{-9} \int \left( T_{\rm R}^{\ast} 
 / \eta_{\rm c} \right) {\rm d}\nu$ [erg~s$^{-1}$cm$^{-2}$sr$^{-1}$] 
 (from Lord \& Young 1990) \\
 $^{\rm d)}$ From Smith (1982) (corrected to include flux $<80~\mu$m using
  IRAS numbers); \\
 \hspace*{1em}in units of the local interstellar radiation field: \\
 \hspace*{1em}$\chi_{\circ} = 
  2\times10^{-4}$ erg~s$^{-1}$cm$^{-2}$sr$^{-1}$ (Draine 1978)}
\end{table*}

\begin{figure*}
\centering
\includegraphics[scale=0.65]{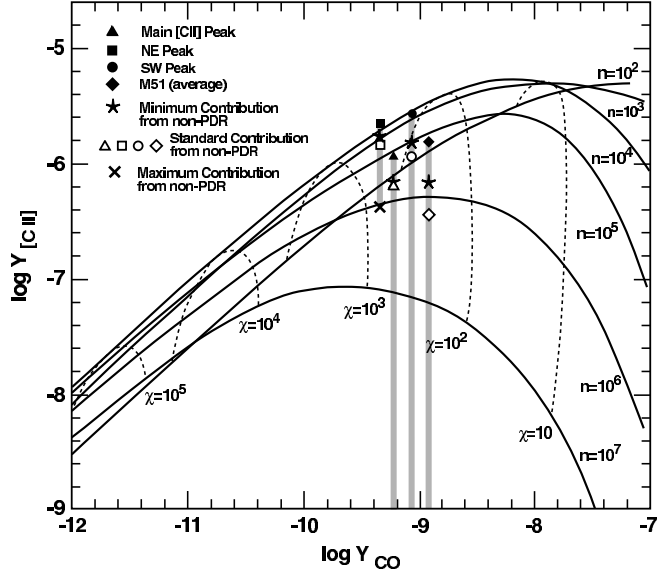}
\caption{
  PDR-plot, showing the density -- FUV-flux phase space of PDRs as a
  function of
  $Y_{\rm [CII]}=I_{\rm [CII]}/\chi_{\rm FIR}$ and
  $Y_{\rm CO}=I_{\rm CO}/\chi_{\rm FIR}$.
  The solid lines and the dashed lines show PDR models at fixed densities 
  and fixed FUV intensities, respectively.
  Peak [C {\small II}] observations and an average over M51 are represented 
  by a filled symbol.
  The location of the filled symbol in the plot marks the phase space for
  the case that the total observed integrated $[$C~II$]$ intensities 
  originate from PDRs.
  The gray bars show the range in phase space that can be covered by
  varying the contribution of $[$C~II$]$ emission originating from PDRs
  for each observed position.
  The stars represent the case that the minium contribution from non-PDRs
  is subtracted form the observed integrated $[$C~II$]$ intensity.
  The open symbols represent the case with the standard contributions from 
  non-PDRs subtracted from the observed $[$C~II$]$ emission and
  the cross represents the case with the maximum contribution from non-PDRs
  subtracted from the observed integrated $[$C~II$]$ intensity.
}
\label{fig6}
\end{figure*}

Figure~\ref{fig6} shows the densities (solid lines) and the FUV
intensities (dashed lines) as functions of
$Y_{\rm [CII]} = I_{\rm [CII]}/\chi_{\rm FIR}$ and
$Y_{\rm CO} = I_{\rm CO}/\chi_{\rm FIR}$ from PDR models (Wolfire,
Hollenbach, \& Tielens 1989, Stacey et al.\ 1991),
where $\chi_{\rm FIR}$ is the integrated FIR intensity expressed 
in units of the local Galactic interstellar radiation field 
$\chi_{\circ} = 2\times10^{-4}$ erg~s$^{-1}$cm$^{-2}$sr$^{-1}$ 
(Draine 1978).
Using the ratios $Y_{\rm [CII]}$ and $Y_{\rm CO}$ eliminate the 
contribution of the beam filling
factor if it assumed that the [C~{\small II}] and $^{12}$CO ($1\to0$) line
and FIR continuum emission are associated with the same PDRs.
The solid lines connect PDR models with the same density while dashed
lines connect PDR models with the same incident FUV intensity.
At a density of about 10$^{4}$~cm$^{-3}$ the $Y_{\rm [CII]}$ ratio reaches
a maximum.
This results in a degeneracy of the PDR model solutions.
If the ($Y_{\rm CO}$, $Y_{\rm [CII]}$) point falls above the 
10$^{2}$~cm$^{-3}$ 
density curve then two different solutions (low and high density) exist for 
the density and the FUV intensity.
The filled symbols represent ``case-1'' (all [C~{\small II}] from
PDRs), the star symbol represent ``case-2'' (minimum contribution
of [C~{\small II}] emission of non-PDRs subtracted) and the open symbols 
represent ``case-3'' (standard contribution of [C~{\small II}]
emission of non-PDRs subtracted) for the individual positions.
The gray bar shows the range of possible [C~{\small II}] contributions
originating in PDRs.
If all the [C~{\small II}] emission would arise from non-PDRs then the
gray bar reaches $Y_{\rm [CII]} = 0$.
The ($Y_{\rm CO}$, $Y_{\rm [CII]}$) positions of
``case-1'' for the NE peak and the SW peak are just at the edge of the 
range for a possible PDR solution but fall clearly within the range of a 
solution if the uncertainty of the [C~{\small II}] measurement is taken 
into account.

The PDR solutions for the main [C~{\small II}] peak, the NE  and SW peak, 
and the total galaxy that we derived from this plot are shown in 
Table~\ref{tab:pdr2}.
As can be seen from the PDR plot the FUV intensities for all the 
positions in M51 are modest.
The beam filling factor which can be determined by the intensity ratio of
the observed FIR radiation and the modeled FUV radiation (Wolfire, 
Hollenbach, \& Tielens 1989, Stacey et al.\ 1991) is also shown in 
Table~\ref{tab:pdr2}.
The calculation of the beam filling factor assumes that about all the FUV 
radiation impinging on the PDR is converted into FIR continuum.
We use this ratio instead of the ratio of the observed to the predicted 
[C~{\small II}] intensity because of the uncertainty of the fraction of
[C~{\small II}] originating from PDRs.
Also the change of the modeled FUV intensities is small for the different
[C~{\small II}] contributions from PDRs.
For the area at the main [C~{\small II}] peak we get a very high beam 
filling factor of nearly 100\% for the low density PDR solution.
In all other cases the beam filling factor is between 7\% and 38\%.

\subsubsection{Mass and column density of the [C~{\small II}] emitting
regions}

We estimate the column density and the mass of the [C~{\small II}]
emitting photodissociated gas using the results of the PDR model above.
For the calculation we assume optically thin emission.
The column density and mass is then given by:
\begin{displaymath}
\begin{array}{rcl}
N_{\rm H}  & = &
 \frac{4 \pi \lambda I_{\rm [CII]}}{h c A X_{\rm C^{+}} \Phi_{\rm b}}
 \left[ \frac{
 \frac{g_{\rm u}}{g_{\rm l}} \exp \left( - \frac{91}{T} \right) }{1 + 
 \frac{g_{\rm u}}{g_{\rm l}} \exp \left( - \frac{91}{T} \right) +
 \frac{n_{\rm cr, H}}{n_{\rm H}}} + \right. \\
& & \left. + \frac{
 \frac{g_{\rm u}}{g_{\rm l}} \exp \left( - \frac{91}{T} \right) }{1 + 
 \frac{g_{\rm u}}{g_{\rm l}} \exp \left( - \frac{91}{T} \right) +
 \frac{n_{\rm cr, e}}{n_{\rm H} x_{\rm e}}} \right] ^{-1} \\
{\rm and} \quad M_{\rm H}  & = & 
N_{\rm H} \frac{ \Omega D^{2} M_{\rm p}}{M_{\odot}} \\
\end{array}
\end{displaymath}
where $\Omega$ is the area in steradian, $M_{\rm p}$ is the proton mass,
$M_{\odot}$ is the mass of the sun, $D$ is the distance,
$h$ is the Planck constant, $c$ is the vacuum speed of light, $A$ is the 
Einstein coefficient for spontaneous emission, $X_{\rm C^{+}} = 
3 \times 10^{-4}$ is the 
abundance of ionized carbon, $g_{\rm u}$ and $g_{\rm l}$ are the 
statistical weights for the upper and lower level, $x_{\rm e} \approx 
3 \times 10^{-4}$ is the ionization fraction, and $n_{\rm cr, H} = 
3.5 \times 10^{3}$ is the mean critical density for collisions with atomic 
and molecular hydrogen.
The density and the beam filling factors were taken from the PDR results
(Table~\ref{tab:pdr2}).
In this calculation we assumed that all carbon is in singly ionized 
form and that the gas temperature is 200~K.
For this gas temperature we assume a critical density for collisions with
electrons of $n_{\rm cr, e} \approx 10$~cm$^{-3}$.
This critical density is deduced from the collision strengths given by 
Blum \& Pradhan (1992).

\begin{table*}
%\begin{center}
\centering
 \caption{Results from the PDR model}
 \label{tab:pdr2}
{\small 
\begin{tabular}{l|cccccccc}
 \hline\hline
 & \multicolumn{2}{c}{Main $[$C~II$]$ peak} & \multicolumn{2}{c}{NE peak} 
 & \multicolumn{2}{c}{SW peak} & \multicolumn{2}{c}{Total M51$^{\rm a)}$} 
 \\ 
 \hline
 & \multicolumn{2}{c}{Density} & \multicolumn{2}{c}{Density} & 
 \multicolumn{2}{c}{Density} & \multicolumn{2}{c}{Density} \\
 & Low & High & Low & High & Low & High & Low & High \\
 & \multicolumn{8}{c}{} \\
 & \multicolumn{8}{c}{From observed values} \\
 $n_{\rm H}$ [cm$^{-3}$] & 300 & $1.2 \times 10^{5}$ & --- & 
 $\approx 10^{4}$ & 900 & $3 \times 10^{3}$ & 200 & $6.3 \times 10^{5}$ \\
 $\chi_{\rm FUV}$$^{\rm b)}$ [$\chi_{\circ}$] & 130 & 460 & --- & 300 & 
 130 & 200& 60 & 200 \\
 $\Phi_{\rm B}$$^{\rm c)}$ & 0.86 & 0.24 & --- & 0.13 & 0.15 & 0.10 & 
 0.38 & 0.12 \\
 & \multicolumn{8}{c}{} \\
 & \multicolumn{8}{c}{Minimum non-PDRs contribution subtracted} \\
 $n_{\rm H}$ [cm$^{-3}$] & $\lesssim 100$ & $4.5 \times 10^{5}$ & 750 & 
 $2 \times 10^{4}$ & 270 & $1 \times 10^{5}$ & --- & $5.4 \times 10^{5}$ \\
 $\chi_{\rm FUV}$$^{\rm b)}$ [$\chi_{\circ}$] & 100 & 520 & 350 & 420 & 
 100 & 300 & --- & 250 \\
 $\Phi_{\rm B}$$^{\rm c)}$ & 1.12 & 0.22 & 0.11 & 0.10 & 0.20 & 0.07 & 
 --- & 0.09 \\
 & \multicolumn{8}{c}{} \\
 & \multicolumn{8}{c}{Standard non-PDRs contribution subtracted} \\
 $n_{\rm H}$ [cm$^{-3}$] & $\lesssim 100$ & $4.6 \times 10^{5}$ & 700 & 
 $3 \times 10^{4}$ & 140 & $1.7 \times 10^{5}$ & --- & $1.4 \times 10^{6}$ 
 \\
 $\chi_{\rm FUV}$$^{\rm b)}$ [$\chi_{\circ}$] & 100 & 530 & 200 & 500 & 80 
 & 300 & --- & 250 \\
 $\Phi_{\rm B}$$^{\rm c)}$ & 1.12 & 0.21 & 0.20 & 0.08 & 0.25 & 0.07 & 
 --- & 0.09 \\
 & \multicolumn{8}{c}{} \\
 & \multicolumn{8}{c}{Maximum non-PDRs contribution subtracted} \\
 $n_{\rm H}$ [cm$^{-3}$] & --- & --- & --- & $1 \times 10^{6}$ & --- & 
 --- & --- & --- \\
 $\chi_{\rm FUV}$$^{\rm b)}$ [$\chi_{\circ}$] & --- & --- & --- & 760 & 
 --- & --- & --- & --- \\
 $\Phi_{\rm B}$$^{\rm c)}$ & --- &  --- & --- & 0.05 & --- & --- & --- & 
 --- \\
\hline
\end{tabular}
}
%\end{center}
\hspace*{2.0cm}\parbox{17cm}{\footnotesize
 $^{\rm a)}$ Averaged over the whole galaxy M51 
 ($\Omega \approx 2.85\times10^{-6}$~sr). \\
 $^{\rm b)}$ In units of the local interstellar radiation field: \\
 \hspace*{1em}$\chi_{\circ} = 2\times10^{-4}$ 
  erg~s$^{-1}$cm$^{-2}$sr$^{-1}$ (Draine 1978) \\
 $^{\rm c)}$ Beam filling factor:
 $\Phi_{\rm B} = \frac{\chi_{\rm FIR}}{\chi_{\rm FUV}}$}
\end{table*}

The results for the column density and the mass are presented in 
Table~\ref{tab:mass}.
For comparison, the molecular masses at NE and SW peak and at the nucleus
within a FIFI beam and of the total galaxy M51
are also determined.
The mass of the molecular gas was derived from the column density of the
molecular hydrogen given in Lord \& Young (1990).
The mass of the [C~{\small II}] 
emitting gas averaged over M51 and at the nucleus is between 2 -- 34\% of 
the molecular mass.
At the NE and the SW peak the fraction of the mass of the 
[C~{\small II}] emitting photodissociated gas is as high as 72\%.

\begin{table*}
%\begin{center}
\centering
 \caption{Column density and mass of the [C II] emitting
  gas compared to the molecular mass}
 \label{tab:mass}
{\footnotesize
\begin{tabular}{r|cccccccc}
 \hline\hline
 & \multicolumn{2}{c}{Main $[$C II$]$ Peak} & \multicolumn{2}{c}{NE Peak} 
 & \multicolumn{2}{c}{SW Peak} & \multicolumn{2}{c}{Total M51$^{\rm a)}$} 
 \\ 
 \hline
 $M_{\rm H_{2}}$$^{\rm b)}$ [$M_{\odot}$] & 
 \multicolumn{2}{c}{$1.0\times10^{9}$} & 
 \multicolumn{2}{c}{$2.6\times10^{8}$} & 
 \multicolumn{2}{c}{$2.5\times10^{8}$} & 
 \multicolumn{2}{c}{$1.35\times10^{10}$} \\
 & \multicolumn{8}{c}{} \\
 & \multicolumn{2}{c}{Density} & \multicolumn{2}{c}{Density}
 & \multicolumn{2}{c}{Density} & \multicolumn{2}{c}{Density} \\
 & Low & High & Low & High & Low & High & Low & High \\
 & \multicolumn{8}{c}{} \\
 \multicolumn{1}{c|}{} & \multicolumn{8}{c}{From Observed Values} \\
 $N_{\rm H}$ [cm$^{-2}$] & $2.4\times10^{21}$ & $7.6\times10^{20}$ &
 --- & $1.5\times10^{21}$ & $2.5\times10^{21}$ & $1.8\times10^{21}$ & 
 $2.2\times10^{21}$ & $4.1\times10^{20}$ \\
 $M_{\rm H}$ [$M_{\odot}$] & $1.5\times10^{8}$ & $4.6\times10^{7}$ &
 --- & $8.9\times10^{7}$ & $1.5\times10^{8}$ & $1.1\times10^{8}$ & 
 $4.6\times10^{9}$ & $8.5\times10^{8}$ \\
 & \multicolumn{8}{c}{} \\
 \multicolumn{1}{c|}{} & \multicolumn{8}{c}{Minimum non-PDR contribution
 subtracted} \\ 
 $N_{\rm HI}$ [cm$^{-2}$] & $3.1\times10^{21}$ & $4.8\times10^{20}$ &
 $4.8\times10^{21}$ & $1.2\times10^{21}$ & $2.7\times10^{21}$ & 
 $6.2\times10^{20}$ & --- & $2.3\times10^{20}$ \\
 $M_{\rm HI}$ [$M_{\odot}$] & $1.9\times10^{8}$ & $2.9\times10^{7}$ & 
 $2.9\times10^{8}$ & $7.4\times10^{7}$ & $1.6\times10^{8}$ & 
 $3.8\times10^{7}$ & --- & $4.9\times10^{8}$ \\
 & \multicolumn{8}{c}{} \\
 \multicolumn{1}{c|}{} & \multicolumn{8}{c}{Standard non-PDR contribution
 subtracted} \\ 
 $N_{\rm HI}$ [cm$^{-2}$] & $2.8\times10^{21}$ & $4.6\times10^{20}$ &
 $2.4\times10^{21}$ & $1.2\times10^{21}$ & $2.9\times10^{21}$ & 
 $4.5\times10^{20}$ & --- & $1.2\times10^{20}$ \\
 $M_{\rm HI}$ [$M_{\odot}$] & $1.7\times10^{8}$ & $2.8\times10^{7}$ & 
 $1.4\times10^{8}$ & $7.4\times10^{7}$ & $1.8\times10^{8}$ & 
 $2.7\times10^{7}$ & --- & $2.6\times10^{8}$ \\
 & \multicolumn{8}{c}{} \\
 \multicolumn{1}{c|}{} & \multicolumn{8}{c}{Maximum non-PDR contribution
 subtracted} \\ 
 $N_{\rm HI}$ [cm$^{-2}$] & --- & --- &
 --- & $4.5\times10^{20}$ & --- & --- & --- & --- \\
 $M_{\rm HI}$ [$M_{\odot}$] & --- & --- & 
 --- & $2.7\times10^{7}$ & --- &  --- & --- & --- \\
 \hline
\end{tabular}
}
%\end{center}
\hspace*{0.1cm}\parbox{17cm}{\footnotesize
 $^{\rm a)}$ Averaged over the whole galaxy M51 
 ($\Omega \approx 2.85\times10^{-6}$~sr).\\
 $^{\rm b)}$ The molecular mass was derived from the H$_{2}$ column 
 densities given in Lord \& Young (1990).}
\end{table*}

\section{Discussion}

\subsection{Starbursts at the NE and SW Peaks}

The [C~{\small II}]/$^{12}$CO(1$\to$0) line intensity ratio is an indicator 
of OB star formation activity in dusty galaxies (Stacey et al.\ 1991).
A ratio of 4400 is typically found in starburst nuclei, while a value near
1200 is more typical of quiesent spiral galaxies.
In the north-east of M51, near the NE peak, we obtain the highest 
[C~{\small II}] to CO integrated intensity ratios in M51 (5000;
for ``case-1'') similar to starburst nuclei.
From their data Lord \& Young (1990) also derived the highest star formation 
efficiency in the north-east region close to the NE peak.
The moderate integrated intensity ratio at the nucleus
($I_{\rm [CII]}/I_{\rm CO} \approx 2000$) indicates moderate star formation 
activity.
Although the peak of the [C~{\small II}] emission is located at the center
of M51 there is also a large amount of CO in this region resulting in the
moderate intensity ratio.
In addition the nucleus of M51 is most likely an AGN 
(Ford et al.\ 1985; Makishima et al.\ 1990; Kohno et al.\
1996; Grillmair et al.\ 1997) and not a starburst region.
The integrated intensity ratio in the south-west of M51 
($I_{\rm [CII]}/I_{\rm CO} \approx 3300$) is smaller than at the NE peak.
Although we can not give a firm value for the fraction of [C~{\small II}]
emission from PDRs in M51 at the NE peak at least 50\% or more of the 
integrated [C~{\small II}] intensity originates most likely from PDRs.
This result and the high [C~{\small II}]/CO(1$\to$0) intensity ratio
suggest that the most active star forming
region in M51 lies in the north-east region, probably near the NE peak.
The maximum of the [C~{\small II}] to CO integrated 
intensity ratio, however, is next to the NE peak between the spiral arms.
This is likely the result of interpolation in the CO map.
No CO data were obtained by Lord \& Young (1990) at the exact location of
the NE peak.
The low [C~{\small II}]/CO(1$\to$0) intensity ratios at the nucleus and the 
SW peak may be the effect of beam dilution.
If the ISM in these regions consists of only a few but highly active star 
forming knots within an extended, less active region the 
[C~{\small II}]/CO(1$\to$0) intensity ratio averaged over the FIFI beam would
be low.
Such a morphology of the ISM is suggested e.g.\ for NGC~4038/39 (Nikola et 
al.\ 1998).

The finding of enhanced star formation activity in the north-east and 
probably also in the south-west is also supported by the distribution of 
the H$\alpha$ emission (van der Hulst et al.\ 1988), the FUV emission 
(Bersier et al.\ 1994), and the $^{12}$CO(2$\to$1) emission 
(Garc\' \i a-Burillo, Gu\'elin, \& Cernicharo 1993),
the FIR continuum (Smith 1982), and the thermal
radio continuum at $\lambda = 2$~cm (Klein, Wielebinski, \& Beck 1984).
All the distributions show emission which either peaks in the north-east
and south-west or only peaks in the north-east but is elongated in 
north-east south-west direction.
However, this distribution is most pronounced in the [C~{\small II}] 
emission with a main focus towards the north-east region.

Additional support for the enhanced star formation activity and the special
location of the activity comes from analysis of the kinematics and dynamics
of M51.
Tully (1974 c) concluded in his analysis that the density-wave pattern in
M51 is confined between the inner Lindblad resonance and corotation and
that the 
spiral arms outside corotation can be accounted for by material clumping 
induced by the galaxy interaction.
Corotation occurs at a radius of about 5.8~kpc (Tully 1974 c; converted to
a distance of M51 of 9.6~Mpc).
Both the NE peak and the SW peak lie at about the same distance from the
center of M51 as corotation.
Thus they are located at the transition zone of the density-wave pattern and 
the spiral arms which are attributed to material clumping.
Especially the SW peak coincides with a position of extreme clumping 
(Tully 1974 c; position ``n'' in Figure~10 therein).
This is also the position where the orbits of all the test particles which 
were originally at or beyond 6.7~kpc pass through (Toomre \& Toomre 1972).
Material clumping can also be seen at the position of the NE peak.
The encounter model of Toomre \& Toomre (1972) shows crowding of test 
particles along the
outer spiral arm which starts at the NE peak where the density-wave arm 
terminates and follows on to north and west to the southern tidal tail.
This spiral arm can be attributed to material clumping (Tully 1974 c).
The relatively weak [C~{\small II}] emission that we observe in the
north-west region might originate from this clumped material.
The coincidence of the location of the NE  and SW peak with the region where
most or all trajectories of the test particles pass through and where the 
density wave arms terminate might indicate that the enhanced [C~{\small II}]
emission is caused by an enhanced mass flow crossing the spiral arms.
This could result in a high rate of cloud-cloud collisions which then 
trigger star formation.
The position of the NE and SW peak at these special locations where the 
tidal tials emerge from the disk in the simulations of Toomre \& Toomre
(1972) also indicates that the star formation there is triggered by the
galaxy interaction.
The trajectories of test particles crossing the spiral arms shown by
Toomre \& Toomre (1972) are almost concentrated in a single spot in the 
south-west in contrast to the north-east where the crossing area
is more spread out.
This would result in a beam dilution effect at the SW peak, resulting in
a smaller [C~{\small II}]/CO(1$\to$0) intensity ratio.

\subsection{Comparison with other Spiral Galaxies}

Except for the NE  and SW peak and the main peak in the center of M51 the
[C~{\small II}] emission is distributed very uniformly over the whole
galaxy.
There is no clear signature of the spiral structure visible in the 
[C~{\small II}] emission.
This could be due to a smoothing effect given our beam size or the presence
of an underlying [C~{\small II}] emission in M51 or both.
However, [C~{\small II}] observations with FIFI of NGC~6946 (Madden et al.\
1993), which is at about the same distance as M51 but a little bit less
extended, shows some spiral structure.
Also the [C~{\small II}] map of M83 taken with FIFI (Geis et al.\ in 
preparation), which has a similar extension as M51 reveals very clumpy
[C~{\small II}] emission following the spiral arms.
This difference in the NGC~6949 and M83 maps and the M51 map supports the
presence of an underlying extended [C~{\small II}] emission in M51.
We have shown in the previous sections that the CNM and the ELDWIM can in 
principle contribute a large fraction to the [C~{\small II}] emission.
The ELDWIM is most likely the main contributer to an underlying 
[C~{\small II}] emission in most of M51.
But in the north-west region of M51 the CNM is most likely the main
source of the [C~{\small II}] emission.
How much the ELDWIM contributes to the [C~{\small II}] emission in M51 
remains unclear.
Even with the use of the [N~{\small II}] 122$\mu$m line it was not 
possible to present a firm statement about the real contribution of the
[C~{\small II}] emission from ionized gas.
Mapping of M51 in the [C~{\small II}] and both [N~{\small II}] lines with 
high spatial resolution might help to disentangle the contributions from
the various ionized gas phases and to distinguish between an on-arm,
inter-arm, and underlying [C~{\small II}] emission.
High spatial resolution would also be required to find out if the PDRs lie 
on a separate spiral arm (maybe between the ionized arm and the molecular
arm or coinciding with the molecular arm) and to distiguish between 
[C~{\small II}] emission from a possible PDR arm and emission from neutral 
gas in the H~{\small I} arm.

The PDRs at the NE  and SW peak might be located in molecular superclouds
of masses $M_{\rm MSC} \approx 10^{7} M_{\odot}$ within 
the giant molecular association with masses of $M_{\rm GMA} \approx 
10^{7} - 10^{8} M_{\odot}$ observed by Vogel, Kulkarni, \& Scoville (1988) 
and Rand \& Kulkarni (1990) in these regions.
This is supported by the low FUV intensity of $\chi_{\rm FUV} \sim
10^{2} \chi_{\circ}$ which we estimated from the PDR models.
Such low FUV intensities are normally expected in PDRs in giant molecular
clouds ($M_{\rm GMC} \approx 10^{5} - 10^{6} M_{\odot}$) (Stacey et al.\
1991, 1993) and might therefore also be the case in PDRs in molecular 
superclouds.

In the nucleus the high density PDR solution is supported by the detection
of the HCN molecule in the center of M51 and in the inner spiral arms 
(Nguyen et al.\ 1992; Kuno et al.\ 1995; Kohno et al.\ 1996).
This finding suggests that the density of the molecular gas in these clouds
is $\ge 10^{5}$~cm$^{-3}$.
No observations of this density tracer are available in the outer
parts of M51 and therefore no statement can be made about which PDR solution
is more likely.

\section{Summary}

We obtained a map of M51 in the [C~{\small II}] fine structure line at 
158~$\mu$m with a spatial resolution of $55''$ (FWHM).
The [C~{\small II}] emission arises from the whole visible extent of M51.
The main peak of the [C~{\small II}] emission is located at the center of
M51 and has an integrated intensity of 
$I_{\rm [CII]} = 1.31 \times 10^{-4}$ erg~s$^{-1}$cm$^{-2}$sr$^{-1}$.
In addition we found a [C~{\small II}] peak in the north-east and in the
south-west of M51.
The NE peak is stronger than the SW peak and has an integrated 
[C~{\small II}] intensity of 70\% of the integrated intensity of the main 
peak.
The NE peak and the SW peak lie roughly on a straight line going through the
center of M51 and they have the same distance to the center as
corotation of the density wave pattern.
Both peaks are located at the position where the density wave pattern 
terminates and where the spiral structure is attributed to material clumping 
due to the interaction of M51 with NGC~5195.
The [C~{\small II}] emission in the NE peak originates mainly
from PDRs.
We also find that the strongest star formation activity occurs in the 
north-east of M51 close to the NE peak.
We suspect that this enhanced star formation activity is triggered by 
cloud-cloud collisions due to a high mass flow crossing the spiral arms.
There might also be high star formation activity in the south-west triggered
by the same effect but which is hidden to our observation due to beam 
dilution.

From the PDR model we derive two possible solutions for the individual 
positions we investigated in M51.
The estimated FUV intensity is similar for both solutions with an
intensity of a few 100 times the local Galactic interstellar radiation 
field.
For the density we obtain very different solutions with $n \sim 10^{2} -
10^{3}$~cm$^{-3}$ for the low density solution and $n \sim 10^{4} - 
10^{6}$~cm$^{-3}$ for the high density solution.
At the center of M51 the high density solution is supported by the detection
of HCN in this region.
The PDRs in the NE  and SW peak are most likely associated with the molecular
superclouds and giant molecular associations found in these regions.
This is supported by the relatively low FUV intensity derived from the
PDR model.
The total mass of the [C~{\small II}] emitting photodissociated regions in 
M51 is $M \approx 2.6 \times 10^{8} M_{\odot}$.
This is only a small fraction (2\%) of the total molecular mass.
If all [C~{\small II}] arises in PDRs that fraction can be as high as 34\%.
In the nucleus the fraction of the [C~{\small II}] emitting 
photodissociated gas ranges between 3\% and 19\% and
at the NE and SW peak the mass of the [C~{\small II}] emitting
gas can be as high as 72\%.

A large fraction of the overall [C~{\small II}] emission in M51 can 
originate in an underlying extended medium.
It remains unclear how much the ELDWIM contributes to this emission, but 
it could be the main source.
The CNM can also be a significant source of [C~{\small II}] emission.
Especially at the north-west part of M51 where most of the
[C~{\small II}] emission might originate from CNM.
At the NE  and SW peak a smaller fraction of the [C~{\small II}] emission
might emerge from CNM.

\acknowledgments                                     
We are grateful to the staff of the KAO for their competent support. 
This work was partially supported by the NASA grant NAG-2-208 to the 
University of California, Berkeley. N.G. was supported in part by a
Feodor-Lynen-fellowship of the Alexander von Humboldt foundation.

\end{document}